\documentclass[%
 reprint,
 amsmath,amssymb,
prl
]{revtex4-2}

\usepackage{graphicx}% Include figure files
\usepackage{dcolumn}% Align table columns on decimal point
\usepackage{bm}% bold math
\usepackage{setspace}
\usepackage{amsmath}
\usepackage{amssymb}
\usepackage{amsthm}
\usepackage{physics}
\usepackage{bbold}
\usepackage{bm}
\begin{document}

\preprint{APS/123-QED}

\title{Direct Numerical Solutions to Stochastic Differential Equations with Multiplicative Noise}% Force line breaks with \\

\author{Ryan T. Grimm}
\author{Joel D. Eaves}%
 \email{Joel.Eaves@colorado.edu}
\affiliation{%
 Department of Chemistry, University of Colorado Boulder, Boulder, CO 80309, USA.
}%

\date{December 9, 2023}% It is always \today, today,

\begin{abstract}
Inspired by path-integral solutions to the quantum relaxation problem, we develop a numerical method to solve classical stochastic differential equations with multiplicative noise that avoids averaging over trajectories. To test the method, we simulate the dynamics of a classical oscillator multiplicatively coupled to non-Markovian noise. When accelerated using tensor factorization techniques, it accurately estimates the transition into the bifurcation regime of the oscillator and outperforms trajectory-averaging simulations with a computational cost that is orders of magnitude lower. 
\end{abstract}

\maketitle

%\tableofcontents

Stochastic differential equations (SDEs) are fundamental models for dynamical systems subjected to statistical fluctuations. In many processes in physics, like Brownian motion, the fluctuations appear as a noisy, additive, external driving force. SDEs of this type characterize the behavior of systems near thermal equilibrium and are some of the most rigorous and well-studied models in statistical physics. 

In many cases, however, the noise is multiplicative and effectively modulates parameters of the system. When the noise is multiplicative, the dynamics of the system become intertwined with the dynamics of the noise, making even approximate solutions challenging. Multiplicative noise occurs in a broad range of systems, including stochastic oscillators \cite{Gitterman2005-nl}, dye lasers \cite{Fox1987-tt}, bi-stable systems that exhibit stochastic resonance \cite{Luo2003-ws}, and active matter \cite{Morozov2010-wx}. 

Kubo was one of the first to formulate approximate theories for multiplicative noise problems in statistical physics, where he modelled how fluctuating magnetic fields give rise to line-broadening in spin resonance experiments \cite{Kubo1962-cq}. He \cite{Kubo1963-iz} also pioneered the use of cumulant expansion methods that van Kampen \cite{Van_Kampen1974-ey, Van_Kampen1974-rm} and many others \cite{Mukamel1978-ch} extended. Cumulant expansions are approximate for multivariate SDEs with multiplicative noise. While low-order cumulant expansions may be tractable, they do not systematically converge and are usually only accurate when the fluctuations are weak, decay quickly, or both. These conditions are sometimes called Markovian. No exact analytical solutions exist for multivariate systems coupled to non-Markovian multiplicative noise sources, and unlike the case of Markovian additive noise, the connection between the SDE and the Fokker-Planck equation remains elusive.   

Open quantum systems present perhaps the most pressing examples of multiplicative noise SDEs, where the noise leads to dephasing and decoherence. Many theories of quantum dissipation model the multiplicative noise as a closed system comprised of a system that is linearly coupled to a heat bath of harmonic oscillators. This formulation removes many uncertainties---for example, the fluctuations satisfy detailed balance. However, because observables require a trace over the heat bath, this approach introduces a many-body problem that is difficult to solve. Numerical methods, like Makri's quasi-adiabatic path integral \cite{Makri1995-ej} (QUAPI) and the hierarchical equations of motion (HEOM) \cite{Tanimura2020-re} have become increasingly popular over the last few decades. They find the reduced density matrix by numerically reducing the computational complexity of the quantum many-body problem, providing insights into quantum relaxation dynamics in regions of parameter space that are otherwise inaccessible.  

In this manuscript, we exploit a correspondence between classical SDEs and open quantum systems to solve SDEs numerically. While our method, which we call ASPEN (\textbf{A}ccelerated \textbf{S}tochastic \textbf{P}ropagator \textbf{E}valuatio\textbf{N}), does not appeal to the path integral, it shares many similarities with QUAPI. Using a stochastic oscillator as a benchmark, we solve for the noise-averaged position directly and compare those results against noise-averaged trajectories. It produces numerically accurate results compared to well-converged trajectory-averages in all regions of the parameter space we have examined. In particular, the method correctly predicts the bifurcation transition \cite{Mallick2006-qa} of the oscillator where some perturbative methods qualitatively fail---similar to the localization-delocalization transition in the analogous spin-boson model---and where trajectory-averaging converges slowly. By accelerating convergence using a tensor-train, we obtain numerically accurate results with orders of magnitude less computational time than trajectory-averaging takes to converge.  

We consider dynamical systems whose equation of motion is isomorphic with the stochastic Liouville equation for the reduced density matrix in an open quantum system \cite{Stockburger2004-sg, Cheng2004-zf}, 
\begin{equation}
    \frac{d\boldsymbol{a}(t)}{dt} = \left[\boldsymbol{L_0} + \boldsymbol{L_1}(t)\right] \boldsymbol{a}(t),
    \label{eq:dyn_sys}
\end{equation}
where $\boldsymbol{a}(t)$ is the state of the system, and $\boldsymbol{L_0}$ is a linear operator that generates the system's dynamics in the absence of fluctuations. The random operator $\boldsymbol{L_1}(t)$ encodes the fluctuations. In classical and quantum mechanics, the objective is to compute the mean state, averaging over the noise $\langle \bm a(t) \rangle$ from a given initial condition $\bm a(t_0)$. The formal solution is $\langle \bm a(t) \rangle = \bm \Phi(t|t_0) \bm a(t_0)$, where $\bm \Phi(t|t_0)$ is Kubo's relaxation operator \cite{Kubo1963-iz}, a positively time-ordered exponential
\begin{equation}
    \boldsymbol{\Phi}(t|t_0) = \left \langle \exp_{\leftarrow}\left((t - t_0)\boldsymbol{L_0} + \int_{t_0}^t ds \, \boldsymbol{L_1}(s) \right) \right \rangle.
\label{eq:propagator}
\end{equation}

One critical difference between SDEs for classical and quantum systems is that in quantum systems, $\bm L_1(t)$ has complex-valued noise fields \cite{Stockburger2004-sg}, which can make trajectory-averaging difficult to implement and slow to converge. One can ignore the imaginary part of the noise correlations, but this violates detailed balance \cite{Cheng2004-zf}. Due to these complexities, accurate solutions are only known for simple open quantum systems, like the spin-boson model, and they are only known in limited regions of parameter space. 

Eq. (\ref{eq:dyn_sys}) embodies many mathematical pathologies of an open quantum system, even when the variables and noise are both real. In particular, although the statistics of $\bm L_1(t)$ may be Gaussian, a cumulant resummation of the relaxation operator does not close at second order other than in the trivial cases where either the noise is Markovian or $\bm L_1(t)$ commutes with itself and with $\bm L_0$ at all times. As a result, there is no way to obtain a simple closed-form expression of the relaxation operator by resumming its cumulants.  

A classical damped harmonic oscillator multiplicatively coupled to a random field is a realization of  Eq. (\ref{eq:dyn_sys}). With $\bm a(t) = ( p(t),x(t) )$ and unit mass, the equation of motion is
\begin{equation}
    \frac{d}{dt} \begin{pmatrix}
p \\
x 
\end{pmatrix} = \left[ \begin{pmatrix}
-\gamma & -\omega_0^2 \\
1 & 0 
\end{pmatrix}
 + \xi(t) \begin{pmatrix}
0 & -1 \\
0 & 0 
\end{pmatrix}
\right ]
\begin{pmatrix}
p \\
x 
\end{pmatrix}.
\label{eq:eom}
\end{equation}
$\bm L_0$ is the first 2x2 matrix on the right hand side and $\bm L_1$ is the second. The decay rate is $\gamma$, $\omega_0$ is the resonance frequency, and $\xi(t)$ is a Gaussian random noise field that obeys an Ornstein-Uhlenbeck process, $d\xi(t) = -\tau_c^{-1} \xi(t) dt + \alpha \sqrt{2 / \tau_c}  \circ dW$, where $dW$ is a Wiener process interpreted in the Stratonovich sense.  While we use this system as a simple model, the parametric stochastic oscillator in Eq. (\ref{eq:eom})  appears in a surprisingly broad set of applications \cite{Gitterman2005-nl} from cosmology \cite{Dodonov2020-pl} to quantum transport phenomena \cite{Tessieri2000-zf}, decoherence \cite{Mukamel1978-ch}, and critical dynamics \cite{Gitterman2004-hp}. 

\begin{figure}[b]
\includegraphics{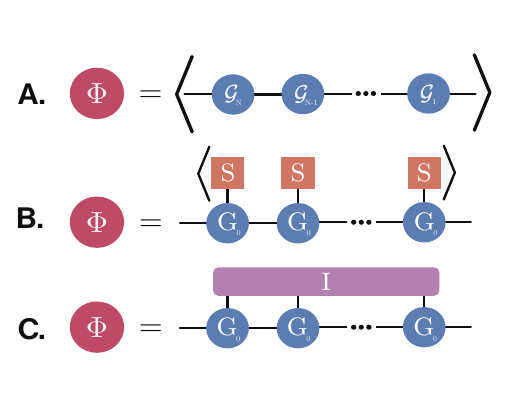}
\caption{\label{fig:method} Schematic derivation of the method. (A) Break the propagation into small steps and apply symmetric Trotter splitting to each. (B) Expand each resulting stochastic matrix using a spectral decomposition to separate the propagator into deterministic matrices $G_0$ and stochastic, scalar exponentials $S$. (C) Compute the average of the scalar part analytically, which generates a high-rank influence tensor $I$ that couples different state-time points.}
\end{figure}

In the weak noise limit or when the noise correlations are short-lived ($\alpha \tau_c \ll 1$), perturbative expansions in $\lambda\bm L_1(t)$ can be accurate and yield useful results, like the Redfield theory for the density matrix in the quantum Liouville equation. We, therefore, repeat the procedure here for Eq. (\ref{eq:eom}) (see SI). Perturbation theories are, in some sense, a natural extension of the scalar cumulant expansion to noncommuting stochastic matrices. As in the quantum literature, however, averaging a time-ordered exponential of a noncommuting operator produces nonunique and ambiguous perturbative results at finite order in $\lambda\bm L_1(t)$. These problems are generic to noncommuting time-dependent stochastic matrices and not specific to quantum relaxation phenomena. 

Mukamel, Oppenheim, and Ross \cite{Mukamel1978-ch} study two different time-ordering prescriptions, called partially ordered products (POP) and chronologically ordered products (COP). When applied to this problem, at second-order in $\lambda$, the COP theory produces a time-nonlocal (TNL) integrodifferential equation for $\langle \bm a(t)\rangle$, while the POP theory yields a time-local (TL) differential equation for it. Formally, the POP convention preserves the stationary Gaussian distribution \cite{Mukamel1978-ch}, while the COP prescription preserves Wick factorization of higher-order moments. In practice, for noncommuting stochastic matrices, these two formal properties give little guidance into which equation of motion will yield the more accurate dynamics for systems that obey Eq \ref{eq:dyn_sys}.  

An alternative theoretical approach takes a direct average of Eq. (\ref{eq:dyn_sys}) and employs the Novokov-Furutsu (NF) theorem \cite{Athanassoulis2019-un} to split moments containing both the system state and noise variable. However, this approach leads to an infinite hierarchy of coupled equations for their correlations \cite{De_Vega2015-xx}. Interestingly, truncating the hierarchy at second-order yields the TNL from perturbation theory. Both approaches, one based on cumulant expansions and another on the application of the NF theorem, yield an infinite number of terms that are not uniformly convergent, making systematic expansions to higher-order fraught. 

The mathematical structure of Eq. (\ref{eq:dyn_sys}) suggests a simplification for calculating the relaxation operator $\bm \Phi$ numerically. Because the stochastic matrix $\bm L_1(t)$ is a constant matrix multiplied by a scalar stochastic variable, $\bm L_1(t) = \xi(t)\bm L_1$, we develop a numerical technique based on the exact cumulant resummation of the scalar stochastic variable $\xi(t)$. Our method, ASPEN, addresses the problem of computing the average of the exponential of noncommuting stochastic matrices. Discretizing the propagation, performing Trotter factorizations, and approximating noncommuting matrices as exponentials of symmetric and antisymmetric matrices allows us to calculate the relaxation operator $\bm \Phi$ by using spectral decomposition and cumulant resummation to average over the scalar noise exactly.

Begin by discretizing the propagation into time steps of duration $\tau$ using the composition property of the propagator, $\boldsymbol{\Phi}(t=N\tau) = \left  \langle \prod_{n = 1}^N \boldsymbol{\mathcal{G}}_n \right \rangle = \left \langle \prod_{n = 1}^N \exp_{\leftarrow} \left ( \tau \boldsymbol{L_0} + (\xi_n^- + \xi_n^+)\boldsymbol{L_1}  \right) \right \rangle,$ where the integration over $\xi(t)$ splits into a lower $\xi_n^- = \int^{\tau (n - 1/2)}_{\tau(n - 1)} ds \, \xi(s)$ and an upper part $\xi_n^+ = \int^{\tau n}_{\tau(n - 1/2)} ds \, \xi(s)$. The first Trotter factorization is a symmetric splitting between the exponential of $\bm L_0$ and $\bm L_1$, $e^{\tau \boldsymbol{L_0} + (\xi_n^- + \xi_n^+)\boldsymbol{L_1} } \approx e^{\tau \bm L_0/2}e^{(\xi_n^- + \xi_n^+)\boldsymbol{L_1}}e^{\tau \bm L_0/2}$. However, unlike in the quantum case, the matrices $\bm L_0$ and  $\boldsymbol{L_1}$ are not Hermitian. They are $2 \times 2$ antisymmetric and defective, respectively. Thus, although $\bm L_1$ has a matrix exponential, it too is defective, and the resulting form is not conducive to performing a cumulant resummation. Expressing $\bm L_1$ as a sum of symmetric and antisymmetric matrices, however, $\boldsymbol{L_1} = -\frac{1}{2}\left(\boldsymbol{\sigma_x} + i \boldsymbol{\sigma_y}\right)$, where $\bm \sigma_x$ and $\bm \sigma_y$ are Pauli matrices, does allow for an optimal approximation of its matrix exponential in terms of matrix products, ultimately allowing a cumulant resummation. Using a splitting approximation for time-ordered exponentials,  $e^{(\xi_n^- + \xi_n^+) \boldsymbol{L_1}} \approx e^{-\frac{1}{2} \xi_n^+ \boldsymbol{\sigma_x}} e^{- \frac{i}{2} (\xi_n^- + \xi_n^+) \boldsymbol{\sigma_y}} e^{-\frac{1}{2} \xi_n^- \boldsymbol{\sigma_x}}$ \cite{Huyghebaert1990-tc}. 

Applying a spectral decomposition to each stochastic matrix ($e^{ (\xi_n^- + \xi_n^+) \boldsymbol{L_1}}$) separates it into a stochastic scalar part and a deterministic matrix product. After averaging the stochastic part over the Gaussian noise, the relaxation operator is
\begin{equation}\boldsymbol{\Phi}(N \tau) = \sum_{ \{\bm \mu \} } \prod_{n = 1}^N \boldsymbol{G}_0(\bm \mu_n) \exp \left ( \sum_{n, m = 1}^N H(\bm \mu_n,\bm \mu_m)  \right),
\label{eq:PI_prop}
\end{equation}
where $\boldsymbol{G}_0(\alpha,\beta,\gamma) =  e^{\tau \boldsymbol{L_0} / 2} \boldsymbol{E}^{\boldsymbol{x}}_{\alpha} \boldsymbol{E}^{\boldsymbol{y}}_{\beta} \boldsymbol{E}^{\boldsymbol{x}}_{\gamma} e^{\tau \boldsymbol{L_0} / 2}$ is a deterministic free propagator. Greek indices correspond to the matrices of the spectral decomposition, for example, $\bm \sigma_x = \sum_{\alpha=\pm 1} \ket{x,\alpha} \alpha \bra{x,\alpha} \equiv \sum_{\alpha=\pm1} \alpha\, \bm E_\alpha^{\bm x}$. For a given time point $n\tau$, there are eight $2\times2$ matrices for $G_0$ corresponding to the entries of the $\bm \mu_n = ( \alpha_n, \beta_n, \gamma_n)$ vector.

The effective pairwise Hamiltonian $H(\bm \mu_n, \bm \mu_m) = \frac{1}{2} \boldsymbol{\lambda}^{\intercal}(\bm \mu_n) \bm C_{n,m} \boldsymbol{\lambda}(\bm \mu_m),$ emerges after applying Gaussian statistics to average over the noise. $\bm C_{n,m} = \langle \boldsymbol{\xi}_n \otimes \boldsymbol{\xi}_m \rangle$ is a correlation matrix,  $\boldsymbol{\lambda}(\bm \mu) = - \frac{1}{2} \begin{pmatrix}
\alpha + i \beta,  
\gamma + i \beta 
\end{pmatrix}$ is a two-component spin with complex-valued components and $\boldsymbol{\xi}_n = \begin{pmatrix}
\xi^-_n,
\xi^+_n \\ 
\end{pmatrix}.$ Written in this way, $\bm \Phi$ has the form of a partition sum, where the sum runs over path variables $\{\bm \mu \} = \{\bm \mu_N, \dots, \bm \mu_1\}$ with $H$ corresponding to a one-dimensional Heisenberg-like Hamiltonian with long-ranged interactions between spins $\bm \lambda(\bm \mu)$. The analogy to equilibrium statistical mechanics is helpful in framing certain results. For example, the interactions are limited to nearest-neighbors in the Markov limit. For the Ornstein-Uhlenbeck process studied here, there is an emergent thermal factor $H \rightarrow \beta H$, where $\beta = \kappa^2$ and $\kappa = \tau_c \alpha$ is the Kubo number. Weak coupling  ($\kappa \ll 1$) therefore corresponds to high temperature. 

Eq. (\ref{eq:PI_prop}) is the central result of this letter. We have obtained a closed form of the relaxation operator, which systematically converges as $\tau \rightarrow 0$. The structure of Eq. (\ref{eq:PI_prop}) is similar to the propagator that one obtains in discretized path integral techniques like QUAPI \cite{Makri1995-ej} for quantum relaxation problems. However, computing $\bm \Phi$ is as computationally complex as calculating the partition function of a one-dimensional quantum spin lattice. Just as the entropy must be extensive, the number of configurations in the partition sum grows exponentially with $N = t/\tau$. Tensor factorization shares some commonalities with transfer matrix methods and is an efficient way of finding the highest weight configurations in the partition sum, taming the curse of dimensionality. The time-evolving matrix product operator (TEMPO) method \cite{Strathearn2018-jc} rephrases the problem of computing the partition sum of a one-dimensional long-ranged quantum lattice into one of contracting higher dimensional tensor objects using a tensor network called a tensor-train. TEMPO has been used to dramatically improve convergence in the QUAPI method. To improve convergence in ASPEN, we borrow the mapping and topology of the TEMPO tensor-train but extend it by adding a node for the propagator $\boldsymbol{G}_0$ and contracting over three legs rather than one for all contractions with the influence tensor (see SI). The structure of our tensor train appears somewhat similar to a recent paper from  Gribben \textit{et al}.  \cite{Gribben2022-vg} for a quantum relaxation problem.

\begin{figure*}
\includegraphics{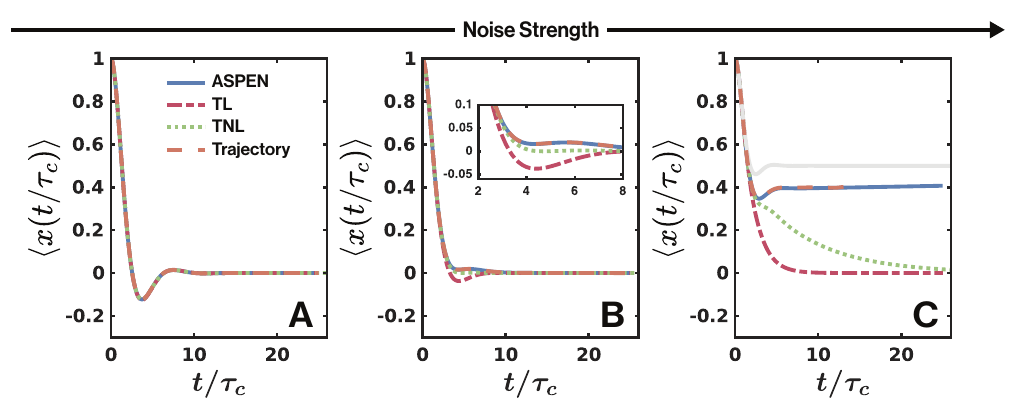}
\caption{\label{fig:osc_bench} Comparing different solution methods for the noise-averaged position of the stochastic parametric oscillator (Eq. \ref{eq:eom}, $\gamma = \omega_0 = 1/\tau_c$).  Exhaustive trajectory averaging (dotted orange line) is arbitrarily accurate and is the standard to which all other methods should be compared. ASPEN agrees quantitatively  (solid blue line) with trajectory averaging (dashed orange) for all noise strengths and all times with orders of magnitude lower computational cost. Time local (TL, red alternating dashed lines) and time non-local (TNL, green dotted line) are the results of the perturbation theories described in the text that correspond to different types of cumulant expansions. (A) At low noise strength $\kappa = 0.5$, all methods agree. (B) At a moderate noise strength $\kappa = 1$, the perturbative methods show moderate disagreement with trajectory averaging and with one another for the transients (inset). (C) At high noise strength, near the bifurcation transition ($\kappa = 1.581$), the oscillator coordinate reaches a nonzero steady state because it cannot dissipate the power of the noise. Perturbation theories fail qualitatively in this regime. The TNL perturbation theory predicts a bifurcation, but at an incorrect Kubo number $\kappa_c = \sqrt{3}$ (shown in grey).}
\end{figure*}
To compare the accuracy of ASPEN and the TL \& TNL perturbation theories, we benchmark them against trajectory averaging, which is computationally expensive but arbitrarily accurate. To average trajectories, we integrated Eq. (\ref{eq:eom}) using an explicit Heun integrator and averaged 350 million trajectories. Initial conditions were $x(0) = 1$, $p(0) = 0$ with initial values of the noise drawn to yield an unconditioned OU process. In the OU process, the noise correlation functions are characterized by a strength $\alpha$ and corrrelation time $\tau_c$ such that correlations obey $\langle \xi(t) \xi(s)\rangle = \alpha^2 e^{-|t - s|/\tau_c}$. In Figure \ref{fig:osc_bench}, we simulate the parametric oscillator defined in Eq. (\ref{eq:eom}), choosing time in units of $\tau_c$ and setting parameters $\omega_0 = \gamma = 1/\tau_c$. For the OU process, the perturbation expansions are equivalent to an asymptotic expansion for the relaxation operator about small Kubo number $\kappa$. For small $\kappa$ ($\kappa =0.5$ in Figure \ref{fig:osc_bench}A), the perturbation theories, both TL and TNL, agree with trajectory averaging and with the ASPEN method quantitatively. As the Kubo number increases to $\kappa = 1$, (Figure \ref{fig:osc_bench}B), however, the TNL and TL begin to deviate with the numerically exact solutions and with one another in the transient regime (inset). At high noise strengths, the parametric stochastic oscillator is known to undergo a noise-induced bifurcation \cite{Mallick2006-qa}, which results in a non-equilibrium steady state where the oscillator decays to a nonzero displacement. In this case (Figure \ref{fig:osc_bench}C), the TL perturbation theory fails qualitatively. While the TNL method predicts the existence and qualitative behavior of the bifurcation, it gives an estimate of the critical Kubo number that is about 10\% higher than found either in trajectory averaging or with ASPEN.

ASPEN yields quantitative agreement with trajectory averaging over all noise strengths and for all times examined. Solutions using ASPEN take no more than an hour on a laptop. While trajectory averaging is arbitrarily accurate, it becomes more computationally costly as the Kubo number increases. Trajectory averaging takes tens of hours of CPU time to achieve similar results to ASPEN. Even with $3.5 \times 10^8$ trajectories, trajectory averaging does not converge for long times ($t/\tau_c \geq 15$) in the bifurcation regime. Thus, ASPEN significantly outperforms trajectory averaging in terms of accuracy and computational cost. 

Drawing on the mathematical isomorphism between classical SDEs and the stochastic Liouville equation of quantum relaxation theory, we have derived a new numerical method for simulating linear dynamical systems containing multiplicative, non-Markovian noise. Because we chose to analyze a stochastic Liouville equation that closely corresponds with the spin-boson model, several results we found have parallels in that system. There, too, the system undergoes a localization-delocalization transition past a critical Kubo number reminiscent of the bifurcation transition in the classical stochastic oscillator. TL and TNL perturbation theories converge at low Kubo number and become unreliable near the bifurcation transition---again, similar to the spin-boson model. 

Methods similar to those that underlie QUAPI and TEMPO---Trotter splitting, spectral decomposition, and tensor-train acceleration---are employed here to solve similar problems and to similar effect. Another similarity ASPEN has with both QUAPI and HEOM is that it becomes much more computationally expensive with increasing Kubo number. In our language, this is because the number of state-time configurations required to compute the relaxation operator is exponentially large. While tensor-train factorization can find the highest-weight states, it is not the only way. Future work will use importance sampling methods, familiar in equilibrium statistical mechanics, to locate and sample the ensemble of the highest weight configurations.

Our work suggests an alternative approach to method development for quantum relaxation phenomena---benchmarking methods not only against quantum problems but against the analogous classical problem. The reason that one devises a numerical solution method for a quantum problem is to probe regions of parameter space that are difficult to explore. There is no guarantee that a numerical method developed in an easy region of parameter space, shown to agree well with other methods developed in the easy region, will perform well in the challenging region. While there are methods for trajectory-averaging complex-valued noise for the quantum Liouville equation, they, too, can be technically challenging to implement and can be slow to converge. Trajectory-averaging the classical system is much more straightforward, arbitrarily accurate, and likely much less computationally demanding---even in challenging regions of parameter space. It is possible, and perhaps even likely, that a method that is accurate in challenging regions of classical parameter space will function well in similarly challenging regions for a quantum system. 
\section*{Acknowledgments} 
R.T.G. was supported by the National Science Foundation Graduate Research Fellowship. This material is based upon work supported by the National Science Foundation Graduate Research Fellowship Program under Grant No. (DGE 2040434). Any opinions, findings, and conclusions or recommendations expressed in this material are those of the author(s) and do not necessarily reflect the views of the National Science Foundation.  This work utilized the Alpine high-performance computing resource  \cite{alpine} at the University of Colorado Boulder. Alpine is jointly funded by the University of Colorado Boulder , the University of Colorado Anschutz, Colorado State University, and the National Science Foundation (award 2201538). 

\bibliography{references}% Produces the bibliography via BibTeX.

\end{document}